\documentclass[referee]{aa}
\usepackage{graphicx}

\begin{document}

\title{A study of the Galactic star forming region IRAS 02593+6016 / S 201 
in infrared and radio wavelengths} 

\author{D.K. Ojha \inst{1,4} \and S.K. Ghosh \inst{1}
\and V.K. Kulkarni \inst{2} \and L. Testi \inst{3} \and R.P. Verma \inst{1}
\and S. Vig \inst{1} }

\offprints{D.K. Ojha, \email{ojha@tifr.res.in}}

\institute{Tata Institute of Fundamental Research, Homi Bhabha Road, Mumbai 
400 005, India \and
National Centre for Radio Astrophysics, Post Bag 3, Ganeshkhind, Pune 411 007,
India \and
Osservatorio Astrofisico di Arcetri, Largo E. Fermi 5, 50125 Firenze, Italy 
\and
National Astronomical Observatory of Japan, Osawa 2-21-1, Mitaka, 
Tokyo 181-8588, Japan
}

\date{Received xxxxxxx; accepted xxxxxxx}

\abstract{
We present infrared and radio continuum observations of S 201 star forming 
region. A massive star cluster is observed in this region, which contains 
different classes of young stellar objects. The near-infrared
colour-colour and colour-magnitude diagrams are studied to discuss the
nature of these sources. We have discovered the knots of molecular 
hydrogen emission at 2.122 $\mu m$ in the central region of S 201. These 
knots are clearly seen along the diffuse emission in north--west 
direction which are probably the obscured Herbig-Haro objects. 
High sensitivity and high resolution radio continuum images from
GMRT observations at 610 and 1280 MHz show an
interesting arc-shaped structure due to the interaction between the 
HII region and the adjacent molecular cloud. The ionization front at the 
interface between the HII region and the molecular cloud is clearly seen 
by comparing the radio, molecular hydrogen and Br$\gamma$ images. The 
emission from the carriers of Unidentified Infrared Bands in the 
mid-infrared 6--9 $\mu m$ (possibly due to PAHs) as extracted from the 
Midcourse Space Experiment survey (at 8, 12, 14 and 21 $\mu m$)
is compared with the radio emission. The HIRES processed IRAS maps at 
12, 25, 60, and 100 $\mu m$, have also been used for comparison.
The spatial distribution of the 
temperature and the optical depth of the warm dust component
around S 201 region, has been generated from the mid-infrared images.
}

\titlerunning{A study of the galactic star forming region S 201.......}
\authorrunning{D.K. Ojha et al.}

\maketitle

\section{Introduction}

The high mass star forming region S 201, corresponding to the radio 
source W 5A, is at a distance of 2.3 kpc from the Sun (Mampaso et al. 1987). 
The S 201 region is coincident with an IRAS source (IRAS 02593+6016) with
increasing spectrum from 12 to 100 $\mu m$. The IRAS fluxes lead to a
luminosity of $\sim$ 1.1$\times$10$^4$ L$_{\odot}$ (Zinchenko et al. 1997)
for this source. The HII region S 201
is part of an extended complex of gas and dust stretched in the east-west
direction and it is bordered on the west by the molecular cloud IC 1848 
identified with the radio source W 5.

Earlier observations of S 201 region in the infrared wavelengths 
(Kleinmann et al. 1979, Thronson et al. 1984, Mampaso et al. 1989, 
Carpenter et al. 1993) and the 
radio (Felli et al. 1987, Fich 1993, Omar et al. 2002) have shown an 
interesting structure presumably due to the interaction between the HII region 
and the adjacent molecular cloud (Martin \& Barrett 1978). The radio source 
has a bright, arc-shaped edge on one side and a smoothly decreasing surface 
brightness distribution on the opposite side as seen in the VLA observations at
2 and 6 cm (Felli et al. 1987). They modeled this configuration in terms 
of a three-dimensional electron distribution produced by the ionizing
radiation of an early-type star located outside a spherical molecular cloud
and found that an O9 zero age main sequence (ZAMS) star placed at a distance of 0.38 pc from a 
molecular cloud of 0.42 pc radius allows a good fit to the their data. 

We have started a near-infrared (NIR) and radio continuum observational 
programme of study of regions of massive star formation. This is our first
paper in this series. In this paper we study the Galactic star forming region
S 201 in the infrared and the radio wavelengths with the aim of identifying the
stellar populations in the directions of the IRAS source 
\mbox{(IRAS 02593+6016)}.
This paper combines new NIR observations from Telescopio 
Infrarosso del Gornergratm (TIRGO) \& Telescopio Nazionale Galileo 
(TNG) telescopes, Italy with mid-infrared (MIR) data 
(8--21 $\mu m$) from Midcourse Space Experiment (MSX) and radio
continuum observations at 610 \& 1280 MHz from Giant Metrewave Radio 
Telescope (GMRT), India.

In Sect. 2, we present the 
broad- and narrow-band NIR observations, as well as complementary
MIR and radio continuum observations. Sect. 3 deals with the results
and the discussion and we summarize our conclusions in Sect. 4.

\section{Observations and data reduction}

\subsection{NIR observations}

\subsubsection{Broad-band images}

The broad-band (JHK) NIR observations were carried out on 23 December 
2000 using the Arcetri NIR camera (ARNICA) mounted on the 1.5m f/20 infrared
telescope TIRGO. ARNICA is equipped with a NICMOS3 256$\times$256 HgCdTe
detector. The plate scale at TIRGO was 0.96\arcsec/pixel and the mean
PSF was approximately 1.5\arcsec--1.8\arcsec (FWHM) during the observations. 

We observed the field in the direction of the IRAS source 02593+6016 
in the three standard J (1.25 $\mu m$, $\Delta\lambda$ = 0.3 $\mu m$), 
H (1.6 $\mu m$, $\Delta\lambda$ = 0.3 $\mu m$), and K (2.2 $\mu m$, 
$\Delta\lambda$ = 0.4 $\mu m$) broad-band filters. A large number of
dithered sky frames were obtained (by shifting the telescope a few arcmin 
off the source in north-west-south-east directions) in all the filters for
sky subtraction and for making flat frames. Total on-target integration 
times were 60 s, 24 s, and 24 s in the J, H, and K bands, respectively. 
The photometric calibration was obtained by observing the standard
stars AS03 \& AS09 (Hunt et al. 1998) in all three bands. Fig. 1 shows the
J, H, and K band images of \mbox{S 201} region. 
The JHK images show the presence of diffuse emission near the
center of the images, with an apparent concentration of embedded stars,
suggesting the presence of a young cluster (see Sect. 3).

Data reduction was done using IRAF\footnote{IRAF is distributed by the 
National Optical Astronomy Observatories, which is operated by the
Association of Universities for Research in Astronomy, Inc. under contract
to the National Science Foundation.} software tasks. All the NIR images went
through standard pipeline procedures like sky-subtraction and flat-fielding.
Accurate photometry was performed on the point 
sources detected in each of the three bands using the DAOPHOT 
(Stetson 1987) routines in IRAF software. Absolute position calibration was 
achieved using the coordinates of a number of stars from the USNO2.0 catalogue.
The completeness limits of the images were evaluated by adding artificial
stars of different magnitudes to the images and determining the fraction
of stars recovered in each magnitude bin. The recovery rate was greater than
90\% for magnitudes brighter than 17, 16 and 15.5 in the J, H and K bands,
respectively. The observations are complete (100\%) to the level of
15.5, 14.5 and 14 magnitudes in J, H and K bands, respectively. The average
photometric error in all colours is $\pm$0.07 mag.

We checked our photometry with the data from 2MASS (The Two Micron All Sky
Survey). The 2MASS All Sky Point Source 
Catalogue\footnote{http://www.ipac.caltech.edu/cgi-bin/gator/nph-dd}
provides J, H \& K$_{\rm s}$ magnitudes for each source. The 2MASS K$_{\rm s}$
filter is centered at 2.17 $\mu m$ and has a bandpass of 0.32 $\mu m$.
Fig. 2 shows the comparison between TIRGO and 2MASS magnitudes in J, H \& K
filters. We see a good linear relation between the two systems, with the
slopes between 0.9--0.97 and the dispersions that increase with the
magnitude.

\subsubsection{Narrow-band images}

The images through narrow-band filters including the molecular hydrogen
(H$_2$ $v$ = 1 -- 0 $S(1)$) transition (2.122 $\mu m$, FWHM = 0.032 $\mu m$),
the Br$\gamma$ line (2.169 $\mu m$, FWHM = 0.035 $\mu m$), and
continuum K (Kcont at 2.275 $\mu m$, FWHM = 0.039 $\mu m$), were obtained
on 13 November, 2002 at 3.58m TNG
telescope at La Palma using the Near-Infrared Camera Spectrometer (NICS).
NICS is the TNG infrared (0.9--2.5 $\mu m$) multimode instrument which is
based on a HgCdTe Hawaii 1024x1024 array (Baffa et al. 2001). The image scale 
was 0.25\arcsec/pixel and the mean PSF was approximately 1.3\arcsec~(FWHM).
Total integration time was of 90 s in all three 
filters. The images were analysed in a similar process as those from TIRGO.
To identify the pure line emission (H$_2$ $v$ = 1 -- 0 $S(1)$ and Br$\gamma$))
one needs to subtract the continuum. That is done by subtracting the K
continuum image from H$_2$ and Br$\gamma$ images  
after aligning and PSF matching.

Fig. 3 shows the H$_2$, Br$\gamma$, H$_2$ continuum subtracted (reveal
the pure H$_2$ line emission), and Br$\gamma$ continuum subtracted images.

\subsection{Mid-infrared data from MSX}

The MSX images in  
A (8.3 $\mu m$), C (12.13 $\mu m$), D (14.65 $\mu m$), and E (21.34 $\mu m$) 
bands (Price et al. 2001) for the region around S 201 
have been used to estimate the
spatial distribution of warm interstellar dust, its temperature and optical
depth. The MSX A and C bands with $\lambda(\Delta\lambda)$ corresponding
to 8.28(3.36) and 12.13(1.72) include several Unidentified Infrared 
emission Bands
(UIBs) at 6.2, 7.7, 8.7, 11.3, and 12.7 $\mu m$. Using a scheme developed by 
Ghosh \& Ojha (2002), the emission in
these UIBs, probably due to the Polycyclic Aromatic Hydrocarbons (PAHs),
have been extracted from the MSX images by correcting for the underlying 
thermal continuum from the interstellar dust. 

\subsection{Mid- and far-infrared data from IRAS}

The data from the IRAS survey in the four bands (12, 25, 60, and 100 $\mu m$)
for the region around S 201 were HIRES processed (Aumann et al. 1990) at
IPAC. The HIRES processed maps in all the four IRAS bands are shown in
Fig. 4. These maps have also been used to generate the maps of dust colour
temperature and optical depth.

\subsection{GMRT Radio continuum observations}

The ionized gas within and around the HII region associated with 
S 201 has been mapped at high angular resolution using the Giant Metrewave
Radio Telescope (GMRT) array. The radio continuum observations have been 
carried out in two frequency bands, viz.,
610 \& 1280 MHz on 04 January, 2002 and 28 September, 2002,
respectively. The sources 3C48 and 3C147 were used as the primary flux 
calibrators for 610 \& 1280 MHz observations, respectively, while the
source 0432+416 was used as a secondary calibrator for both the observations. 
The GMRT antennas and their configurations are discussed in detail by
Swarup et al. (1991).

Data reduction was done in classic AIPS. Bad data (dead antennas, interference,
spikes, etc.) were identified and flagged using UVFLG \& TVFLG. Images of the
field were formed by Fourier inversion and cleaning (IMAGR). The initial 
images were improved by self-calibration (CALIB) in both phase and 
amplitude.  

Figs. 5 \& 6 show the radio continuum images of S 201 region generated from
the GMRT observations at 1280 \& 610 MHz respectively. 
The radio continuum images have a resolution of 
4.5\arcsec$\times$2.5\arcsec at 1280 MHz, and 8.7\arcsec$\times$5.7\arcsec
at 610 MHz. The images have an rms of 42 $\mu$Jy beam$^{-1}$ at 1280 MHz, and
47 $\mu$Jy beam$^{-1}$ at 610 MHz. The total flux densities are 0.68 Jy \&
0.77 Jy at 1280 MHz \& 610 MHz, respectively.

\section{Results and discussion}

\subsection{The embedded cluster}

Inside a $\sim$ 3.5\arcmin$\times$3.4\arcmin ~field centered on IRAS 02593+6016
source, 114 stars are found to be common to all the three JHK bands and 153 
stars are common to the HK bands alone, with magnitude errors less than 0.2. 
Fig. 7 shows a colour-colour (CC) diagram for the 114 stars detected in the JHK
bands. The solid and broken heavy curves represent the unreddened main 
sequence dwarfs and giant branch
(Koornneef 1983) and the parallel dashed lines are the reddening vectors for
early and late type stars (drawn from
the base and tip of the two branches) that encloses reddeded main 
sequence objects. The dotted line indicates the locus of T-Tauri
stars (Meyer et al. 1997). We have assumed that 
A$_{\rm J}$/A$_{\rm V}$ = 0.282; A$_{\rm H}$/A$_{\rm V}$ = 0.175 and
A$_{\rm K}$$_{\rm s}$/A$_{\rm V}$ = 0.112 (Rieke \& Lebofsky 1985).

Most of the sources have colours of reddened 
photospheres but some of the stars lying outside the region of reddened main
sequence objects (right of the reddening line for early type stars) are 
mostly young stellar
objects (YSOs) with intrinsic colour excess. By de-reddening the stars (on the
CC diagram) that fell within the reddening vectors encompassing the main
sequence stars, we found visual extinction (A$_V$) towards each star. The 
individual extinction values range from 0 to 18 magnitudes
with the average
foreground extinction of A$_V$ $\sim$ 5 mag. The stars lying on the left
side of the reddening band are mostly foreground stars as supported by
their low values of A$_V$.

About 12\% of the sources (within 0.43 pc radius around the position of 
the IRAS source) detected in J, H \& K bands show an infrared 
excess (H-K $>$ 1, J-H $>$ 1, A$_V$ $\simeq$ 8--14). These YSOs are shown
as star symbols in Fig. 7 and they are concentrated close to the center of 
the embedded young stellar cluster. 
The most massive O6--O8 type star is located S-W of the cluster and is 
surrounded by at 
least two massive stars of spectral types earlier than B2, both of them
showing an infrared excess (see Fig. 8). 

Fig. 8 shows the H-K vs K colour-magnitude (CM) diagram for all the sources 
detected in HK bands. The vertical solid lines (from the left to the right) 
represent the main sequence curve reddened by 0, 20 and 40 magnitudes,
respectively. We have assumed a distance of 2.3 kpc to the source to 
reproduce the main sequence data on this plot. 
The horizontal slanting lines in Fig. 8 trace the reddening zones for each
spectral type. However, it should be noted that the spectral 
types inferred from the CM diagram are only upper limits when
the stars present infrared excess.
YSOs with an infrared excess found from CC diagram (Fig. 7) are shown as
star symbols. However, it is important to note that even those stars not
shown with a star symbol may represent YSOs with intrinsic colour
excess, since these stars detected in the H and K bands are not detected
in the J 
band due to their very red colours. The three massive and luminous 
stars discussed earlier are located at the top of the CM diagram   
(K $<$ 11.8).

Following the method described in Testi et al. (1999) we estimated 
the number of stars in the central regions of the cluster, within
40\arcsec~(0.4 pc), from the most massive star in the cluster (\#2).
We found an excess of $\sim$ 38 objects above the background with 
K band magnitude less than 15.5, corresponding to our 90\% completeness
level. The corresponding stellar volume density in the inner region 
of the cluster would then be $\sim$ 10$^2$ stars/pc$^3$, on the low side 
compared to
the typical densities of objects around early Herbig Be stars in the 
survey of Testi et al. (1999). However, to make a sensible comparison
we need to estimate the lowest mass object that can be probed with our
observations, which can be estimated following the method outlined
in Testi et al. (1998). Assuming the distance of 2.3 kpc, an age in 
the range 0.5--1 Myr, and an extinction in K band between 0 and 1 mag (up to
A$_{V}$ $\sim$ 10), the derived magnitude limit corresponds to 
M $\sim$ 0.5 M$_\odot$ (using Palla \& Stahler (1999) pre-main sequence 
evolutionary tracks)), which is significantly higher than most of the limits 
in Testi et al. (1999). 

\subsection{Comparison of infrared and radio continuum observations}

The radio maps from GMRT at 1280 and 610 MHz (Figs. 5 \& 6) 
display a striking cometary morphology with sharp boundaries toward the N--NE,
and diffuse emission extending toward S--SW (an angular diameter of
approximately $\sim$ 4\arcmin). This kind of morphology is
produced by the winds and shocks from the massive stars in the vicinity
of the molecular clouds, which are expected to compress the nearmost edges
of nearby clouds and develop the cometary structure
(Bachiller et al. 2002). Three bright 
near-infrared sources are present within the radio nebulosity. One of the 
sources (id\#1 in Fig. 1) with NIR excess (H-K $>$ 1)
is very close to the radio peak suggesting the spectral type of B1
in the H-K vs K CM diagram. This source is also coincident, within errors, with
the H$_2$O maser position present in the region (Blair et al. 1980).
The brightest infrared source (id\#2; K=10.24, H-K = 1.40, J-H = 2.66) 
deeply embedded in the cloud (A$_V$ $\sim$ 22), is located 
$\sim$ 26\arcsec~($\sim$ 0.29 pc) west of the radio peak. The projected
distance of source \#2 from the cloud is approximately 33\arcsec~(0.37 pc). 
The colour and IR luminosity of this source indicate that it is an O6--O8 type 
star. The position of \#2 coincides with the ``secondary peak'' detected
by Felli et al. (1987) in the radio and also seen in the GMRT radio images
(see Figs. 1 \& 5). 

It is interesting to revisit the model of Felli et al. (1987) after
the results of our new near-infrared observations. Their
modeling of the VLA observations was based on the assumption of a single
star providing all the ionizing photons, while our observations clearly
reveal a rich cluster within the radio continuum emission.
The estimated spectral types from CM diagram and the total Lyman continuum 
photons (using the table in Panagia 1973) for the five bright stars within the
young cluster are shown in Table 1. 
The most massive star (\#2) alone provides more than 90\% of the total
Lyman continuum supply rate, explaining the success of the model based on
a single source of the ionizing radiation. It is also interesting to 
note that our observations confirm the suggestion by Felli et al. (1987)
of the co-location of the massive star responsible for the ionization
of the nebula and the secondary peak of radio emission.

The quadrupole transition ($v$ = 1 -- 0 $S(1)$) of molecular hydrogen at
2.122 $\mu m$ is an excellent tracer of shock emission
and photo-dissociation regions (PDRs). In particular, in PDRs the 
molecular hydrogen emission traces the first neutral layer beyond the
ionization front, a typical, well studied case is that of the PDR at the
interface between the Orion Nebula and the Orion molecular cloud (the
Orion Bar, see Walmsley et al. 2000). In fact, the model of
Felli et al. (1987) implicitly predicts that a PDR should be
located at the interface between the ionized and molecular gas, to the 
east of the arc-shaped feature in the radio continuum maps.
Our near-infrared narrow band images fully support this picture, as
we detect diffuse H$_2$ emission where the PDR region should be
located (see Fig. 9). Additionally, in the north-east corner of the image 
we detect an H$_2$ knot reminiscent of Herbig-Haro objects.


The continuum subtracted Br$\gamma$ image displays a morphology very similar 
to the radio continuum maps, with a bright, arc-shaped edge on one side and a 
smoothly decreasing surface brightness distribution on the opposite side
and a peak coinciding exactly with the radio continuum one (Figs. 3 \& 10). 
The similar morphology between the Br$\gamma$ and radio continuum maps
indicates that there are no steep extinction gradients across the S 201
region.

The ionization front at the interface between the HII region and the 
molecular cloud is clearly seen by comparing the radio, molecular
hydrogen and Br$\gamma$ images (Figs. 3, 5 \& 6). A more detailed analysis 
from the GMRT radio observations will be presented in a future paper.   

\begin{table}
\caption[]{Bright stars within the young cluster}
\begin{tabular}{c c c c c c c c}
\hline
RA (B1950)  & DEC (B1950) & J     & H     & K     & Sp. Type   & Lyman Continuum Photons$^1$ & Id. \\
hh:mm:ss    & dd:mm:ss    & mag   & mag   & mag   & (from CM)  & (Photons/sec) & (Figs. 1, 7 \& 8)   \\
\hline
02:55:17.72 & +60:04:01.2 & 13.28 & 12.23 & 11.57 & B2.5 & 2.5$\times$10$^{44}$\\
02:55:21.45 & +60:04:27.8 & 13.67 & 11.94 & 10.70 & B0 & 2.3$\times$10$^{47}$ & \#3\\
02:55:22.65 & +60:03:59.2 & 16.17 & 14.14 & 12.99 & B4 & -- & \\
02:55:23.74 & +60:04:09.2 & 14.30 & 11.64 & 10.24 & O6--O8 & 5.7$\times$10$^{48}$ & \#2 \\
02:55:27.28 & +60:04:12.6 & 14.80 & 13.15 & 11.79 & B1 & 1.9$\times$10$^{45}$ & \#1\\
\hline
\end{tabular}
\noindent
$^1$ From Panagia (1973)
\end{table}

\subsection{Spatial distribution of Unidentified Infrared emission Bands 
(UIBs) around \mbox{S 201} region}

The MSX surveyed the entire Galactic plane within $|b|\le5^\circ$ 
in four mid-infrared spectral bands centered at 8.28, 12.13, 14.65 and
21.34 $\mu m$, with image resolution of 20\arcsec~(Price et al. 2001).
We have used the scheme developed by Ghosh \& Ojha (2002) to extract the
contribution of Unidentified Infrared emission Bands (UIBs) from the 
mid-infrared MSX images of S 201 region in the four bands. The scheme models
the observations with a combination of thermal emission (gray body) from
interstellar dust and the UIB emission from the gas component, under 
reasonable assumptions. The spatial distribution of emission in the 
UIBs with an anugular resolution $\sim$ 20\arcsec~(intrinsic to MSX survey)
has been extracted and is shown in Fig. 11. 

The UIBs emission map has been compared with
the GMRT observations at 1280 MHz in Fig. 12. The positions of the peaks
and other morphology (along N-S to E-W) compare rather well between the 
UIBs and the radio emission maps though we see a strong extended component of
UIBs emission along S-E as compared to the radio emission.

\subsection{Dust optical depth and temperature maps}

\subsubsection{From MSX}

The MSX maps were used to obtain maps of warm dust temperature and 
optical depth (Ghosh \& Ojha 2002). 
Since the range of frequencies covered by the MSX bands is limited, we
assume a power law dependence of the dust emissivity on frequency 
of the form $\epsilon_{\lambda} \propto \lambda^{-1}$.
The optical depth and dust temperature maps have been presented in 
Fig. 13.

The positions of the peaks of the optical depth and temperature are different.
The optical depth map is morphologically similar to the UIBs intensity map
and the peaks in the two coincide indicating presence of the high densities 
near the embedded sources. However, the temperature map shows a more extended
distribution with decreasing temperature gradient toward S to S-W. The 
temperature map peaks close to the boundary of UIBs emission toward west
(48\arcsec~or 0.5 pc from the UIBs emission peak).


\subsubsection{From HIRES}

We have used HIRES maps to generate maps of the dust colour temperature 
($T(12/25)$ \& $T(60/100)$), and optical depth ($\tau_{12}$ \& $\tau_{100}$) 
around S 201 region. The intensity maps at 12, 25, 60 and 100 $\mu m$ (Fig. 4) 
were spatially averaged before computing $T(12/25)$, $\tau_{12}$, $T(60/100)$, 
and $\tau_{100}$ in a manner similar to that described by Ghosh et al. (1993) 
for an emissivity law of $\epsilon_{\lambda} \propto \lambda^{-1}$. The dust
optical depth and temperature maps are presented in Fig. 14.

\subsubsection{Comparison between MSX and HIRES maps}

A comparison of the $\tau_{10}$ maps generated from the higher 
angular resolution MSX maps (Fig. 13a) and that based on IRAS HIRES maps at 
12 and 25 $\mu m$ (the best resolution among the 4 bands) is in order.
The latter has been presented in Fig. 14a, which is scaled to 10 $\mu m$ by 
$\lambda^{-1}$ emissivity law to make the comparison with MSX $\tau_{10}$
map. The peak optical depth and the effective FWHM for the central maximum are 
2.13 $\times 10^{-4}$ and 54\arcsec~respectively for the map based on MSX. 
The corresponding values from the IRAS--HIRES maps are 6.6 $\times 10^{-5}$ 
and 60\arcsec. 

These derived values are in reasonable agreement 
considering the fact that they are based on instruments with very different 
angular resolutions. The difference in the peak values of $\tau_{10}$ may be 
a result of two effects, viz., beam dilution and a clumpy interstellar medium.


The peak dust column density implied by the optical depth map at 100 $\mu m$ 
(Fig. 14b), is much higher than that implied by the $\tau_{10}$
map from MSX data (Fig. 13a). This is not surprizing, since the former has 
main contribution from the cooler component of dust grains (T $<$ 80 K)
lying in the larger outer envelope of the cloud. Only the dust 
grains very close to the exciting source can contribute to $\tau_{10}$
map.

Compared to MSX, the HIRES maps at 60 and 100 $\mu m$ trace the distribution 
of colder dust (35--55 K). However, the hot spot due west and a cooler
plateau due east can be seen in the two maps (Figs. 13b \& 14c), 
which shows the similar sense of temperatures between the MSX and HIRES
(T(12/25)). 




\section{Conclusions}

A detailed radio and infrared study of an embedded young stellar cluster 
associated with the Galactic star forming region S 201 is presented here.
The colour-colour and colour-magnitude diagrams have been constructed to
identify young stellar objects and estimate their spectral 
types. The high sensitivity and high angular resolution radio continuum maps based
on the GMRT observations at 610 and 1280 MHz have been generated.
These radio maps show interesting morphological details, including
an interesting arc-shaped structure highlighting the interaction between the 
HII region and the adjacent molecular cloud. Three luminous infrared sources
have been identified which are located within the radio nebulosity. 
The spatial distribution
of temperature and optical depth of the interstellar dust component 
in the S 201 region, has been presented based on the mid- and far-infrared
measurements from the MSX and IRAS (HIRES) missions.

We have detected a compact embedded star cluster located within the ionized
nebula responsible for the radio continuum emission.
The cluster has richness similar to the clusters surrounding
early type Herbig Be stars; the most luminous member is consistent
with an O6--O8 zero age main sequence star. Our analysis confirms that
this single star is responsible for most of the Lyman continuum emission
required to sustain the ionized radio nebula, with all the other
cluster members contributing at most 10\% of the ionizing photons.
This explains the reason for the success of the model of Felli et al. (1987)
(which assumed a single massive star for excitation),
in explaining the radio continuum morphology of S 201.
Our observations also confirm the position of this massive star as well as
the location of the molecular cloud being eroded by the ionizing radiation.
The PDR at the interface between the ionized gas and the molecular cloud 
is traced to the west of the bright arc-shaped feature in radio continuum and 
Br$\gamma$, by
the diffuse H$_2$ emission detected in our narrow-band
near-infrared images.

\begin{acknowledgements}

We thank the staff of the GMRT that made the radio observations possible. The 
GMRT is run by the National Centre for Radio Astrophysics of the Tata Institute
of Fundamental Research. It is a pleasure to thank the Arcetri and
TNG technical staff and the TNG operators for their assistance during the 
NICS observing runs. We also thank the TIRGO team, especially Filippo Mannucci,
for nice scheduling and service observing at the Gornergrat.

This paper is based on observations made with the Italian Telescopio
Nazionale Galileo (TNG) operated on the island of La Palma by
the Centro Galileo Galilei of the CNAA (Consorzio Nazionale per
l'Astronomia e l'Astrofisica) at the Spanish Observatorio del
Roque de los Muchachos of the Instituto de Astrofisica de Canarias.

This publication makes use of data products from the Two Micron All Sky
Survey, which is a joint project of the University of Massachusetts and the
Infrared Processing and Analysis Center/California Institute of Technology,
funded by the National Aeronautics and Space Administration and the National
Science Foundation. IPAC is thanked for providing HIRES processed IRAS
data. 

This research made use of data products from the
Midcourse Space Experiment, the processing of which was funded by the
Ballistic Missile Defense Organization with additional support from from NASA
Office of Space Science. This research has also made use of the NASA/IPAC
Infrared Science Archive, which is operated by the Jet Propulsion Laboratory,
California Institute of Technology, under contract with National Aeronautics
and Space Administration.  

DKO was supported by the JSPS (Japan) through a fellowship during which 
some part of this work was done. We thank Francesco Palla for providing
us with their PMS grids.

\end{acknowledgements}

\newpage
\begin{figure}
\caption{TIRGO J, H \& K band images (clockwise from the top left) of S 201
region. North is the top and east to the left. The total integration times
are 60 s in J, 24 s in H, and 24 s in K bands. The abscissa and the ordinates
are in B1950.0 epoch. The bottom left figure shows the TIRGO K band image
overlayed by GMRT radio contours (see \S 3.2 for discussion).}
\label{fig1}
\end{figure}

\begin{figure}
\centering
\caption{Comparison between TIRGO and 2MASS magnitudes in J, H \& K bands.
The continuous lines show the linear fit to the magnitudes. The slopes to
the linear fits range from 0.9 to 0.97 between the two systems in the three
filters.}
\label{fig2}
\end{figure}

\begin{figure}
\caption{H$_2$, Br$\gamma$, continuum subtracted Br$\gamma$, and
continuum subtracted H$_2$ line images of the central region
($\sim$ 88\arcsec$\times$88\arcsec) of S 201 (clockwise from the top
left). Notice also the presence of H$_2$ knots near the center and upper
left corner of the image (bottom left figure).
North is the top and east to the left. The total integration time is
90 s in all the filters.
The abscissa and the ordinates are in B1950.0 epoch.}
\label{fig3}
\end{figure}

\begin{figure}
\caption{The HIRES processed IRAS maps for S 201 region in the four bands
(clockwise from the top left) -- (a) 12 $\mu m$ with peak = 21 Jy/sq. arcmin,
(b) 25 $\mu m$ with peak = 122 Jy/sq. arcmin, (c) 60 $\mu m$ with peak =
622 Jy/sq. arcmin, and (d) 100 $\mu m$ with peak = 415 Jy/sq. arcmin.
The isophot contour levels in 12 $\mu m$ are 95, 90, 80, 70, 60, 50, 40, 30,
20 \& 10 \%, in 25 \& 60 $\mu m$ are 95, 90, 80, 70, 60, 50, 40, 30, 20, 10,
5 \& 2.5 \%, and in 100 $\mu m$ are 95, 90, 80, 70, 60, 50, 40, 30, 20, 10 \&
5 \% of the respective peaks. The abscissa and the ordinates are in B1950.0
epoch.}
\label{fig4}
\end{figure}

\begin{figure}
\centering
\caption{GMRT high resolution map of S 201 at 1280 MHz. The resolution is
4.5$\times$2.5 arcsec$^2$ along PA = -23$^\circ$, and the rms noise in the
map is 42 $\mu$Jy beam$^{-1}$.}
\label{fig5}
\end{figure}

\begin{figure}
\centering
\caption{GMRT high resolution map of S 201 at 610 MHz. The resolution is
8.7$\times$5.7 arcsec$^2$ along PA = -14$^\circ$, and the rms noise in the
map is 47 $\mu$Jy beam$^{-1}$.}
\label{fig6}
\end{figure}

\begin{figure}
\centering
\caption{Colour-Colour diagram for the 114 sources detected in JHK bands in
S 201 region.
The locii of the main sequence and giants branch are shown by the solid
curve (dwarf) and broken heavy curve (giants) taken from Koornneef (1983).
The three parallel dashed straight lines follow the reddening vectors taken
from Rieke \& Lebofsky (1985). Crosses on the dashed lines are separated
by A$_V$ = 5 mag. The dotted line represents the locus of T-Tauri stars
(Meyer et al. 1997). The star symbols represent YSOs with intrinsic colour
excesses within 0.43 pc radius around the center of the cluster. Three
bright stars within the cluster are represented by \#1, \#2 \& \#3
(see Fig. 1 \& \S 3).}
\label{fig7}
\end{figure}

\begin{figure}
\centering
\caption{Colour-Magnitude diagram for the 153 sources detected in HK bands.
Stars represent the YSOs identified from Fig. 7. The vertical solid lines
from left to right indicate the track of main sequence dwarfs at 2.3 kpc
reddened by 0, 20 and 40 magnitudes, respectively. The intrinsic colours are
taken from Koornneef (1983). Slanting horizontal lines identify the reddening
vectors (Rieke \& Lebofsky 1985). The symbols are same as shown in Fig. 7.}
\label{fig8}
\end{figure}

\begin{figure}
\centering
\caption{Continuum-subtracted molecular hydrogen emission at 2.122 $\mu m$
around S 201 region. The contour plot shows the 1280 MHz radio continuum
emission from GMRT. Notice also the presence of knots near the center
and upper left corner of the
image. The abscissa and the ordinates are in B1950.0 epoch.}
\label{fig9}
\end{figure}

\begin{figure}
\centering
\caption{Continuum-subtracted Br$\gamma$ image of S 201 region. The
contour plot shows the 1280 MHz radio continuum emission from GMRT.
The abscissa and the ordinates are in B1950.0 epoch.}
\label{fig10}
\end{figure}

\begin{figure}
\centering
\caption{The spatial distribution of total radiation in Unidentified Infrared
emission Bands (UIBs) for the region around S 201, as extracted from the MSX
images.
Contour levels are drawn at 5, 10, 20, 30, 40, 50, 60, 70, 80, 90, 95 and
99\% of the peak intensity of 4.11$\times$10$^{-5}$ $W m^{-2} sr^{-1}$.
The abscissa and the ordinates are in B1950.0 epoch.}
\label{fig11}
\end{figure}

\begin{figure}
\centering
\caption{The emission in UIBs (grey scale) in the mid-infrared (6--9 $\mu m$).
The contours show the radio emission at 1280 MHz as measured by GMRT.
The abscissa and the ordinates are in B1950.0 epoch.}
\label{fig12}
\end{figure}

\begin{figure}
\caption{(a) The spatial distribution of dust optical depth ($\tau_{10}$ at
10 $\mu m$) for the S 201 region, as extracted from the MSX images. The
contour levels are at
$\tau_{10}$ = 0.1$\times$10$^{-4}$, 0.2$\times$10$^{-4}$, 0.4$\times$10$^{-4}$,
0.8$\times$10$^{-4}$, 1.2$\times$10$^{-4}$, 1.6$\times$10$^{-4}$ \&
2.0$\times$10$^{-4}$. The peak value in this map is 2.13$\times$10$^{-4}$.
(b) The spatial distribution of dust temperature and the contours correspond
to the temperatures 130, 120, 110, 90, 70, 50 \& 30 K. The abscissa and the
ordinates are in B1950.0 epoch.}
\label{fig13}
\end{figure}

\begin{figure}
\caption{(a) The dust optical depth ($\tau_{10}$) distribution from the HIRES
12 and 25 $\mu m$ maps, scaled to 10 $\mu m$ by $\lambda^{-1}$ emissivity
law. The contours represent 95, 90, 80, 70, 60, 50, 40, 30, 20, 10, 5 \& 2\%
of the global peak value of 6.63$\times$10$^{-5}$. (b) The dust optical depth
($\tau_{100}$) distribution from the HIRES 60 and 100 $\mu m$ maps
for the region around S 201 assuming a dust emissivity law of
$\epsilon_{\lambda} \propto \lambda^{-1}$. The contours represent
95, 90, 80, 70, 60, 50, 40 \& 30 \% of the global peak value of
3.05$\times$10$^{-3}$. (c) The dust temperature ($T(12/25)$) distribution and
the contours are drawn from 140, 145 to 215 K in step of 10 K (from left to
right in the figure). The peak value in this map is 230 K and occurs
at the extreme right position (RA = 02:58:56.4, DEC=+60:15:38) of the map.
(d) The dust temperature ($T(60/100)$) distribution and the contours
correspond to the temperatures 55, 50, 45, 40 \& 37 K.
The abscissa and the ordinates are in B1950.0 epoch.}
\label{fig14}
\end{figure}

\end{document}